\documentclass[prb,reprint,amsmath,amssymb,floatfix,superscriptaddress]{revtex4-1}
\usepackage{graphicx}
\usepackage{relsize}
\usepackage{hyperref}  
\usepackage{enumerate}
\usepackage{xcolor}
\usepackage{adjustbox}
\renewcommand{\vec}[1]{\mathbf{#1}}

\begin{document}

\title{Spin-orbit coupling and spin-triplet pairing symmetry in $\mathbf{Sr_2 Ru O_4}$}

\author{Zhiqiang Wang}
\affiliation{Department of Physics and Astronomy, McMaster University, Hamilton, Ontario, L8S 4M1, Canada}
\affiliation{James Franck Institute, University of Chicago, Chicago, Illinois, 60637, USA}
\author{Xin Wang}
\affiliation{Department of Physics and Astronomy, McMaster University, Hamilton, Ontario, L8S 4M1, Canada}
\author{Catherine Kallin}
\affiliation{Department of Physics and Astronomy, McMaster University, Hamilton, Ontario, L8S 4M1, Canada}
\date{\today}

\begin{abstract}
Spin-orbit coupling (SOC) plays a crucial role in determining the spin structure of an odd parity psedospin-triplet Cooper pairing state.
Here, we present a thorough study of how SOC lifts the degeneracy among different
p-wave pseudospin-triplet pairing states in a widely used microscopic model for $\mathrm{Sr_2 Ru O_4}$, combining a Ginzburg-Landau (GL) free energy expansion, a symmetry analysis of the model, and numerical weak-coupling renormalization group (RG) and random phase approximation (RPA) calculations. These analyses are then used to critically re-examine previous numerical results on the stability of chiral p-wave pairing. The symmetry analysis can serve as a guide for future studies, especially numerical calculations, on the pairing instability in $\mathrm{Sr_2 Ru O_4}$ and can be useful for studying other multi-band spin-triplet superconductors where SOC plays an important role. 
\end{abstract}


\maketitle

\section{Introduction}
Understanding an unconventional superconductor requires identifying and understanding both its superconducting order parameter symmetry and the pairing mechanism. The two
are intimately connected. In $\mathrm{Sr_2RuO_4}$, both of these are still not well understood. Early experiments, including muon spin relaxation~\cite{Luke1998}, NMR~\cite{Ishida1998},
Polar Kerr effect~\cite{Xia2006} measurements, point toward a spin-triplet chiral p-wave pairing~\cite{Mackenzie2003,Kallin2016}, which is a two-dimensional (2D) analog of the A-phase of
Helium 3~\cite{Leggett1975} and is potentially useful for topological quantum computing~\cite{Read2000,Nayak2008}.

However, it is difficult to reconcile the spin-triplet chiral p-wave picture with several other experiments~\cite{Mackenzie2017}. Chiral edge currents have been predicted for the chiral p-wave
pairing state but not detected~\cite{Matsumoto1999,Kirtley2007}; splitting of the superconducting transition temperature $T_c$ in the presence of an in-plane magnetic field
or a uniaxial strain~\cite{Hicks2014,Steppke2017} is expected but not found. Recent NMR experiments~\cite{Pustogow2019,Ishida2019} report a significant drop of the spin susceptibility in the superconducting
phase measured in an in-plane magnetic field, which contradicts previous measurements~\cite{Ishida1998} and suggest either spin-triplet helical or singlet pairing,
although strong spin-orbit coupling~\cite{Haverkort2008,Tamai2019} can complicate the interpretation of the experimental data.

Most theoretical studies~\cite{Scaffidi2014,Roising2018} on the pairing mechanism are connected to spin or charge fluctuation mediated superconductivity, inspired by work on Helium-3~\cite{Leggett1975}.
However, spin fluctuations in $\mathrm{Sr_2RuO_4}$ are complicated due to the multi-orbital nature of its normal state. The normal state of $\mathrm{Sr_2RuO_4}$
contains two quasi-1D $\alpha$ and $\beta$ bands, derived mainly from the Ru $t_{2g}$ $d_{xz}$, $d_{yz}$ orbitals, and one quasi-2D band from the $d_{xy}$ orbital.
Although early on it was proposed that the superconductivity is dominated by one set of the three bands~\cite{Agterberg1997}, more recent calculations suggest that
superconductivity on the three bands is comparable and indicate that the three orbitals should be treated simultaneously. A further complication in a microscopic analysis
comes from the sizable spin-orbit coupling (SOC) which entangles the three orbital degrees of freedom
with spin. The effect of SOC on the normal state Fermi surface (FS) has been emphasized previously in Ref.~\onlinecite{Haverkort2008} and was recently found to be larger than
previously thought~\cite{Tamai2019}. However, the effect of SOC on the superconducting state is still poorly understood.

Understanding the effect of SOC on the superconducting phase is crucial to address the relative stability of chiral p-wave and 
helical p-wave pairing states. This is because in the absence of SOC, and in the weak-coupling limit, all spin-triplet p-wave pairing states
are degenerate due to the unbroken spin rotation symmetry~\cite{Sigrist1999}. A mechanism to lift the degeneracy in the absence of SOC
is to consider the spin fluctuation feedback effect due to the superconducting condensate itself, which spontaneously breaks the spin rotation symmetry and
modifies the pairing interaction. This mechanism
is responsible for the stability of the Helium-3 A phase~\cite{Leggett1975} and has been used to stabilize the chiral state in theories of $\mathrm{Sr_2RuO_4}$. However,
in a Ginzburg-Landau free energy expansion in terms of the superconducting order parameter near $T_c$, the feedback effect only appears at fourth-order; while the SOC effect
can split $T_{c}$ of different spin triplet states at quadratic order~\cite{Sigrist1999}. Therefore, it is important to understand how the normal state SOC affects the stability
of different pairing states.

The effect of SOC on the spin triplet pairing states in $\mathrm{Sr_2RuO_4}$ has been studied previously in
Refs.~\onlinecite{Sigrist1999,Ng2000,Ng2000b,Yanase2003,Annett2006,Raghu2010,Puetter2012,Yanase2014} semi-analytically to various degrees and included in different
numerical calculations~\cite{Scaffidi2014,Zhang2018,Romer2019,Roising2019}, using different models and approaches. However,  a systematic and more complete treatment is lacking. Also, conflicting statements have been made regarding the degeneracy among different p-wave pairing
states in the presence of SOC. In this paper, we present a complete Ginzburg-Landau free energy analysis of the SOC effect on the superconducting state at quadratic
order in the order parameter. Then we focus on a 2D three-band microscopic model with SOC and identify the terms that lift the degeneracy among different
p-wave states based on a symmetry analysis of the model. The results are supplemented with numerical weak-coupling RG and RPA calculations~\cite{Raghu2010,Scaffidi2014}.
This model has been adopted in different numerical calculations~\cite{Yanase2003,Puetter2012,Raghu2010,Wang2013,Scaffidi2014,Roising2019,Zhang2018,Wang2019,Romer2019} under different approximations to determine the dominant
pairing instability for $\mathrm{Sr_2RuO_4}$. Our analysis shows that some of the previous numerical results obtained in certain parameter regimes are incorrect. Since our
results are obtained largely based on symmetries of the model, they also apply beyond weak-coupling and provide a guide to future numerical calculations.  Furthermore, some of the conclusions and analysis here can be applied to other multi-band spin triplet superconductors, where SOC is important for the pairing.

The rest of the paper is organzied as follows. 
In Sec.~\ref{sec:GL}, a complete GL analysis of SOC effects on triplet states is presented.
In Sec.~\ref{sec:microscopic}, we study the SOC induced GL free energy terms for $\mathrm{Sr_2RuO_4}$ based on a widely studied 2D three-band microscopic model
using analytical symmetry analyses and numerical weak-coupling RG calculations.
In Sec.~\ref{sec:SRO}, we reexamine the chiral p-wave instability in $\mathrm{Sr_2RuO_4}$, where we provide a new phase diagram calculated within the RPA for the
microscopic model, and also generalize the 2D analysis to 3D.
Sec.~\ref{sec:conclusion} contains our conclusions.
Some details of the derivations are relegated to Appendices, including details on the extension of this work to 3D models of $\mathrm{Sr_2RuO_4}$.

\section{General Ginzburg-Landau analysis} \label{sec:GL}
In the presence of SOC, spin is not a good quantum number. However, time reversal and inversion symmetries still ensure a two-fold degeneracy at each $\vec{k}$ point
in the Brillouin zone, which can be used to define a pseudospin
and to classify all possible pairing states into pseudospin singlet and triplet sectors.
Here, we focus on pseudospin triplet $p$-wave pairing states.

For a general pseudospin triplet state the order parameter is a $2\times 2$ matrix,
\begin{gather} \label{eq:OPdef}
\hat{\Delta}(\vec{k}) \equiv \sum_{\mu=\{x,y,z\}} \sum_{j=\{x,y\}} d_{j}^\mu \;  \sigma_\mu i \sigma_y \; \psi_j(\vec{k}).
\end{gather}
where $\sigma_{\mu}$ are Pauli matrices in pseudospin space; $\psi_j(\vec{k})$ are two basis functions in $\vec{k}$-space that transform like
$k_x$ and $k_y$ under the $D_{4h}$ point group.

In the absence of SOC, the GL free energy at quadratic order in the superconducting order parameter is
\begin{subequations}
\begin{align}
f_{2}^0  & =  \alpha^0(T) \; \bigg \langle \frac{1}{2}\mathrm{Tr}[\hat{\Delta}^\dagger(\vec{k}) \hat{\Delta}(\vec{k}) ] \bigg\rangle_{\mathrm{FS}} \label{eq:f20} \\
& = \alpha^0(T) \; \sum_{\mu=\{x,y,z\}}\sum_{j=\{x,y\}} |d_{j}^\mu|^2,
\end{align}
\end{subequations}
where the superscript `0' indicates quantities defined for zero SOC. $\alpha^0(T) \propto (T_c^0 - T)$ and $\langle \cdots \rangle_{\mathrm{FS}}$ means averaged over the FS.
The trace, $\mathrm{Tr}[\cdots]$, is performed in pseudospin space.

In general, the presence of SOC breaks both the full pseudospin $SU(2)$ rotation and spatial $D_{4h}$ symmetries. The remaining symmetry group for a 2D model
of $\mathrm{Sr_2RuO_4}$ is $D_{4h}^{\hat{L}+\hat{S}}
\otimes U(1)^{C}$, where $D_{4h}^{\hat{L}+\hat{S}}$ is the $D_{4h}$ point group whose symmetry operations act simultaneously on the spatial $\vec{k}$ and pseudospin spaces.
$U(1)^{C}$ is the charge $U(1)$ gauge symmetry. Time-reversal and inversion symmetries are also assumed, although they might be spontaneously broken in the ground state. To derive the most general form of the GL free energy terms at
quadratic order we consider all possible contractions of $(d_i^\mu)^* d_j^\nu$, viewed as a rank-4 tensor, such that the contracted results are a scalar that is invariant under
all symmetry operations of $D_{4h}^{\hat{L}+\hat{S}} \otimes U(1)^{C}$. This leads to five terms in the GL free energy,
which are tabulated in Table.~\ref{tab:SOCcomplete}. Details of the derivation can be found in Appendix~\ref{app:SOCDerivation}. 
\begin{table}[htp]
  \centering
    \caption{All possible SOC induced GL free energy terms at quadratic order in $\hat{\Delta}$ for pseudospin triplet pairing states of a 2D model.
    For 3D models there are additional terms, which can be found in Appendix~\ref{app:SOC3D}.  }
  \begin{adjustbox}{width=0.95\linewidth}
      \begin{tabular}{@{\hskip 0.2in} c @{\hskip 0.2in} c  @{\hskip 0.2in}}
  \hline
  \hline
     GL terms   &  Expressions in terms of $d^\mu_j$  \\
  \hline
 $f^{\mathrm{SOC}, 1}_2$ \qquad  &  \qquad $|d^z_x|^2+|d^z_y|^2$  \qquad    \\
  \hline
  $f^{\mathrm{SOC}, 2}_2$ \qquad & \qquad $(d_x^x)^* d_y^y + (d^y_y)^* d_x^x$  \qquad   \\
  \hline
 $f^{\mathrm{SOC}, 3}_2$ \qquad & \qquad $(d_y^x)^* d^y_x + (d_x^y)^* d_y^x$   \qquad    \\
  \hline
 $f^{\mathrm{SOC}, 4}_2$ \qquad &  \qquad $|d_x^x|^2 + |d_y^y|^2$ \qquad      \\
  \hline
 $f^{\mathrm{SOC}, 5}_2$ \qquad & \qquad $|d_x^y|^2 +|d_y^x|^2$  \qquad       \\
  \hline
  \end{tabular}
    \end{adjustbox}
  \label{tab:SOCcomplete}
\end{table}

We can also write the SOC induced terms in terms of $\hat{\Delta}$. When the pseudospin rotation symmetry is broken, order
parameter products other than $\hat{\Delta}^\dagger \hat{\Delta}$, such as $\hat{\Delta}^\dagger \sigma_i \Delta$ and $\hat{\Delta}^\dagger \sigma_i \Delta \sigma_j$, can also appear in Eq.~\eqref{eq:f20}~\cite{Vollhardt1990}.
Considering all such combinations that are invariant under the symmetry group $D_{4h}^{\hat{L}+\hat{S}}\otimes U(1)^{C}$ leads
to the same conclusion that there are five independent terms in the GL free energy at quadratic order. The results can be found in Table.~\ref{tab:SOCcomplete2}
of Appendix~\ref{app:SOCDerivation}.

Some of the terms in Table~\ref{tab:SOCcomplete} have been identified previously,\cite{Sigrist1999,Ng2000,Ng2000b,Yanase2003,Annett2006,Raghu2010,Yanase2014}
 but Table~\ref{tab:SOCcomplete} provides the most complete form of all possible SOC induced terms at quadratic order. 
 These terms in general lift the degeneracy among different p-wave states, which belong to the five irreducible representations of the $D_{4h}$ group and are classified in Table~\ref{tab:D4h}.
 Depending on the symmetries
of microscopic models, some of these terms may or may not appear. In the following, we focus on a particular 2D three-band interaction
model~\cite{Scaffidi2014}, identify the SOC induced terms, and analyze how they affect the relative stability of different p-wave pairing states.

\begin{table}[htp]
  \centering
    \caption{Irreducible representations (irrep.) of the $D_{4h}^{\hat{L}+\hat{S}}$ point group. The order parameters are given for 2D models. Only the pseudospin triplet $p$-wave pairing states are considered. 
    The first four irrep., $\{A_{1u}, A_{2u}, B_{1u}, B_{2u}\}$, give helical pairings that do not break  time reversal symmetry; while the $E_{u}$ irrep. supports two chiral states,
    $\hat{z} (k_x \pm i k_y)$, that spontaneously break time reversal symmetry. }
  \begin{tabular}{@{\hskip 0.5in} c @{\hskip 0.5in} l  @{\hskip 0.5in}}
  \hline\hline
   irrep. &  Order parameter  \\
 \hline
   $A_{1u}$ &  $\vec{d}(\vec{k})= \hat{x} k_x + \hat{y} k_y $   \\

   $A_{2u}$ &  $\vec{d}(\vec{k}) = \hat{x} k_y - \hat{y} k_x $  \\

   $B_{1u}$ &  $\vec{d}(\vec{k}) = \hat{x} k_x - \hat{y} k_y $ \\

   $B_{2u}$ &  $\vec{d} (\vec{k}) = \hat{x} k_y + \hat{y} k_x $  \\

  $E_{u}$ &  $\vec{d} (\vec{k}) = \hat{z} (k_x, \, k_y) $  \\
\hline
\hline
  \end{tabular}
  \label{tab:D4h}
\end{table}

\section{Microscopic determination of the SOC induced terms} \label{sec:microscopic}
We consider the microscopic model Hamiltonian, 
\begin{gather}
H= H_K + V,
\end{gather}
where $H_K$ is the kinetic energy part that gives rise to the normal state Fermi surfaces, and $V$ is the interaction.
In addition to hopping terms, $H_K$ contains a SOC term, which, written in $\vec{k}$ space, is
\begin{gather}
2 \eta \vec{L}\cdot \vec{S}= \eta \sum_{\ell,m,n=1,2,3} i \epsilon_{\ell m n} c^\dagger_{\vec{k},m,s} \sigma^\ell_{s s^\prime} c_{\vec{k},n , s^\prime},
\end{gather}
where $\{1,2,3\}=\{d_{yz},d_{xz},d_{xy}\}$ orbitals, and $\{s,s^\prime\}$ are the actual spins, not the pseudospins to be defined below.
$\epsilon_{\ell m n}$ is the fully anti-symmetric tensor and $\eta$ is the SOC strength. $c^\dagger$ (c) is the electron creation (annihilation) operator. 

Following Ref.~\onlinecite{Scaffidi2014} we write $H_K$ in the basis $\Psi(\vec{k})=[c_{\vec{k},1,\uparrow};c_{\vec{k},2,\uparrow};c_{\vec{k},3,\downarrow};
c_{\vec{k},1,\downarrow}; c_{\vec{k},2,\downarrow};c_{\vec{k},3,\uparrow}]^T$, such that it is block diagonal
\begin{gather} \label{eq:Hblock}
H_K(\vec{k})=
\begin{pmatrix}
H_{\uparrow\uparrow}(\vec{k}) & 0 \\
0 & H_{\downarrow\downarrow}(\vec{k})
\end{pmatrix},
\end{gather}
where
\begin{gather}
H_{s s} (\vec{k})=
\begin{pmatrix}
\epsilon_{yz}(\vec{k})            & g(\vec{k}) + i s \eta           & -s \eta \\
g(\vec{k}) - i s \eta     & \epsilon_{xz}(\vec{k})                     & i \eta \\
- s \eta                  & - i \eta                          & \epsilon_{xy}(\vec{k})
\end{pmatrix}. \label{eq:HK}
\end{gather}
$\epsilon_{yz},\epsilon_{xz}$ and $\epsilon_{xy}$ describe intra-orbital hoppings; while $g(\vec{k})$ is the only inter-orbital hopping for a 2D model.

The interaction~\cite{Scaffidi2014} we consider is a multi-orbital on-site Kanamori-Hubbard type interaction
\begin{align} \label{eq:Kanamori2}
V& = \frac{U}{2} \sum_{i,a} n_{i,a,\uparrow} n_{i,a,\downarrow} + \frac{U^\prime}{2} \sum_{i,a\ne b,s,s^\prime} n_{i,a,s} n_{i,b,s^\prime}  \nonumber \\
& + \frac{J}{2} \sum_{i,a\ne b, s,s^\prime} c^\dagger_{i a s } c^\dagger_{i b s^\prime} c_{i a s^\prime} c_{i b s} \nonumber \\
& + \frac{J^\prime}{2} \sum_{i, a\ne b, s\ne s^\prime} c^\dagger_{i a  s} c^\dagger_{i a s^\prime}  c_{i b s^\prime}  c_{i b s}.
\end{align}
$n_{i,a,s}\equiv c^{\dagger}_{i,a,s} c_{i,a,s}$ is the spin and orbital resolved electron density operator at site $i$. $U$ ($U^\prime$)
is the intra-orbital (inter-orbital) repulsive Hubbard interaction. $J$ is the Hund's coupling, and $J^\prime$ the pair hopping. 
 The Hund's coupling term can be also written as~\cite{Georges2013}
$ - J \sum_{i,a\ne b} (\vec{S}_{i,a}\cdot \vec{S}_{i,b} + n_{i,a} n_{i,b}/4)$,
where $\vec{S}_{i,a}$ is the orbital resolved electron spin vector operator at site $i$ and $n_{i,a}=n_{i,a,\uparrow} + n_{i,a,\downarrow}$. 
The Kanamori-Hubbard interaction $V$ is derived from the Coulomb interaction and is invariant under $SO(3)$ rotations in the $t_{2g}$ d-orbital space,
provided $J^\prime=J$ and $U^\prime=U-2J$~\cite{Georges2013}.  Crystal field splitting in $\mathrm{Sr_2RuO_4}$ in general
lowers the symmetry of the interaction in the orbital space, which, however, does not affect our following discussions.
Each of the four terms of $V$ is $SU(2)$ spin rotational invariant.
The repulsive $V$ can give rise to Cooper pairing instabilities in non-s wave channels~\cite{Kohn1965}.

\subsection{Hamiltonian in the pseudospin basis} \label{sec:pseudospin}
Using $a_{i, a, \sigma}^\dagger$ ($a_{i, a, \sigma}$) for electron creation (annihilation) operators with the pseudospin $\sigma$ and orbital $a$ at site $i$ we define
\begin{gather}
(a_{i,1,\sigma}^\dagger , a_{i, 2, \sigma}^\dagger , a_{i, 3, \sigma}^\dagger ) \equiv (c_{i,1, \sigma}^\dagger, c_{i, 2, \sigma}^\dagger,  c_{i, 3, \bar{\sigma}}^\dagger),
\end{gather}
where $\bar{\sigma}=\downarrow (\uparrow)$ if $\sigma=\uparrow (\downarrow)$. 
Written in the pseudospin basis, $\widetilde{\Psi}(\vec{k})=[a_{\vec{k},1,\uparrow}; a_{\vec{k},2,\uparrow};  a_{\vec{k},3,\uparrow};
a_{\vec{k},1,\downarrow}; a_{\vec{k},2,\downarrow}; a_{\vec{k},3,\downarrow}]^T$, the kinetic energy part $H_K(\vec{k})$
remains the same as in Eq.~\eqref{eq:HK}, whose $H_{\uparrow\uparrow}$ ($H_{\downarrow\downarrow}$) block can be identified with pseudospin $\uparrow$ ($\downarrow$).

Rewriting the interaction $V$ in Eq.~\eqref{eq:Kanamori2} in terms of $\{a^\dagger,a\}$ and denoting the new interaction by $\widetilde{V}$, we have
\begin{gather} \label{eq:Kanamori3}
\widetilde{V} = \widetilde{V}_{U} + \widetilde{V}_{U^\prime}+\widetilde{V}_J +\widetilde{V}_{J^\prime},
\end{gather}
where
\begin{subequations}
\begin{align} \label{eq:Kanamori-new}
\widetilde{V}_{U}               =  &   \frac{U}{2} \sum_{i,a} n_{i,a,\uparrow} n_{i,a,\downarrow}  \\
\widetilde{V}_{U^\prime}   =  & \frac{U^\prime-J/2}{2} \sum_{i,a\ne b, \sigma, \sigma^\prime} n_{i,a,\sigma} n_{i,b,\sigma^\prime},   \\
\widetilde{V}_J                  =  & -J \sum_i \bigg\{ \sum_{a\ne b} \vec{S}_{i,a}\cdot \vec{S}_{i,b} \nonumber \\
                                              &   -2 \sum_{a=\{1,2\}} \big[ S^y_{i,a} S^y_{i,3} + S^z_{i,a} S^z_{i,3}\big] \bigg\}, \\
\widetilde{V}_{J^\prime}    = & \frac{J^\prime}{2} \sum_{i,\sigma \ne \sigma^\prime}\bigg\{ \sum_{ a\ne b=\{1,2\}} -  \sum_{ a\ne b=\{2,3\}} - \sum_{a\ne b=\{1,3\}} \bigg\} \nonumber \\
                                            &   a^\dagger_{i a  \sigma} a^\dagger_{i a \sigma^\prime}  a_{i b \sigma^\prime}  a_{i b \sigma}.
\end{align}
\end{subequations}
In these equations, all operators are in terms of $\{a^\dagger,a\}$: $n_{i,a,\uparrow} = a^\dagger_{i, a, \uparrow} a_{i, a, \uparrow}$, etc. In the following, we identify
the terms in the Hamiltonian $H_K + \widetilde{V}$ that breaks the pseudospin rotational symmetry.

\subsection{Degeneracy at $g(\vec{k})=J=J^\prime=0$} \label{sec:degeneracy}
Although the presence of $H_{\mathrm{SOC}}$ breaks spin rotation symmetry in the normal state, it does not necessarily lead to a symmetry breaking
in the pseudospin space and, therefore, the degeneracy among different pseudospin triplet p-wave pairing states may remain intact. In the current model,
this is the case when both $g(\vec{k})\equiv 0$ and $J=J^\prime=0$. This has been
pointed out previously in Ref.~\onlinecite{Yanase2003} by a direct expansion of the effective interaction in the Cooper pairing channel
in terms of the SOC constant $\eta$ up to quadratic order. Here, we provide a proof purely based on symmetry.

First notice that $H_K$ can be brought into a pseudospin $SU(2)$ invariant form
by the following unitary transformation (written in the $\vec{k}$ space)
\begin{gather}\label{eq:U}
\mathcal{U}: \{ a^\dagger_{\vec{k}, 1, \downarrow}, a_{\vec{k}, 1, \downarrow}\} \rightarrow \{- a^\dagger_{\vec{k}, 1, \downarrow},- a_{\vec{k}, 1, \downarrow}\},
\end{gather}
if there is no inter-orbital hopping term, i. e., $g(\vec{k})\equiv 0$ in Eq.~\eqref{eq:HK}.
In this case, under the $\mathcal{U}$ transformation,
\begin{gather}\label{eq:HKtilde}
\widetilde{H}_K \equiv \mathcal{U}^\dagger H_K \mathcal{U} = H_{\uparrow\uparrow} \otimes \sigma_0,
\end{gather}
where $\sigma_0$ is the identity matrix in the pseudospin space. 

When $J=J^\prime=0$, the $\mathcal{U}$ transformation leaves $\widetilde{V}$ in Eq.~\eqref{eq:Kanamori3} unchanged,  which is
pseudospin $SU(2)$ rotational invariant since $\widetilde{V}$ and $V$ share the same form. 
Therefore, if both $g(\vec{k})\equiv 0$
and $J=J^\prime=0$, the whole microscopic Hamiltonian after the $\mathcal{U}$ transformation, 
\begin{gather}
\widetilde{H}=\widetilde{H}_K + \widetilde{V},
\end{gather}
is pseudospin $SU(2)$ invariant.
Consequently, all p-wave pseudospin triplet pairing states resulting from the microscopic Hamiltonian are degenerate. 
This conclusion does not depend on how the microscopic model is treated, i. e., whether the pairing states
are calculated in weak-coupling RG~\cite{Scaffidi2014}, RPA~\cite{Zhang2018,Romer2019} or other methods.

\subsection{SOC induced terms due to finite $g(\vec{k})$ but with $J=J^\prime=0$} \label{sec:SOCtppp}
When $g(\vec{k})\ne 0$, after the $\mathcal{U}$ transformation, the kinetic energy part of the Hamiltonian can be written as $\widetilde{H}_K+\delta \widetilde{H}_K$
 with $\widetilde{H}_K$ given in Eq.~\eqref{eq:HKtilde} and
\begin{gather} \label{eq:deltaHK}
\delta \widetilde{H}_K (\vec{k}) \equiv  2 g(\vec{k}) \{ S^z_{12}(\vec{k}) + h.c.\},
\end{gather}
where $S^z_{12}(\vec{k})\equiv 1/2 \sum_{\sigma,\sigma^\prime}  a^\dagger_{\vec{k}, 1, \sigma} \sigma^z_{\sigma,\sigma^\prime} a_{\vec{k}, 2, \sigma^\prime}$
is the inter-orbital pseudospin operator along the $z$-direction. For $g(\vec{k})$, to be specific, we consider the nearest neighbor inter-orbital
hybridization as in Ref.~\onlinecite{Scaffidi2014}, $g(\vec{k})=-4 t^{\prime\prime\prime} \sin k_x \sin k_y$, where $t^{\prime\prime\prime}$ is the corresponding hopping integral.

Clearly, $\delta \widetilde{H}_K (\vec{k})$ breaks the full pseudospin rotational symmetry. 
It contributes to the GL free energy a term which, to the first order in $t^{\prime\prime\prime}/t$, is 
\begin{gather} \label{eq:deltaFtppp}
\delta \mathcal{F}= \langle \delta \widetilde{H}_K (\vec{k}) \rangle =  a_2 \; \big[ f_2^{\mathrm{SOC},2} -f_{2}^{\mathrm{SOC},3} \big].
\end{gather}
$a_2 \propto t^{\prime\prime\prime}/t$, and the expressions of $f_2^{\mathrm{SOC},2}$ and $f_{2}^{\mathrm{SOC},3}$ are given in Table~\ref{tab:SOCcomplete}. The average $\langle \cdots \rangle$ is performed in a mean-field p-wave pairing state obtained at $t^{\prime\prime\prime}=0$ and over the $\vec{k}$ space.
In arriving at this equation we have used: (1) because of the $\sin k_x \sin k_y$ dependence in $g(\vec{k})$, only $(d_x^\mu)^* d_y^\nu$ type terms can appear in $\delta \mathcal{F}$ so that $\langle \cdots \rangle$ does not vanish after the $\vec{k}$ average; (2) $\delta \widetilde{H}_K (\vec{k})$ has a remaining symmetry in the pseudospin
space; it is invariant under pseudospin rotations about the $z$-axis.
Written in terms of the components of the $\hat{\Delta}$ matrix, $\delta \mathcal{F}=a_2 (-i/2)
\big\{ \Delta_{\uparrow\uparrow,x} \Delta_{\uparrow\uparrow,y}^* -\Delta_{\downarrow\downarrow,x} \Delta_{\downarrow\downarrow,y}^* - c.c.\big\}$.
The subscript `$x$' indicates that the quantity transforms as $k_x$ under the spatial $D_{4h}$ group. This term has been identified in Refs.~\onlinecite{Yanase2003,Yanase2014}
using a quite different approach. Our derivation makes the microscopic symmetry origin of the term manifest.

\begin{figure}[htp]
\centering
\includegraphics[width=\linewidth,trim={22mm 0mm 0mm 0mm},clip]{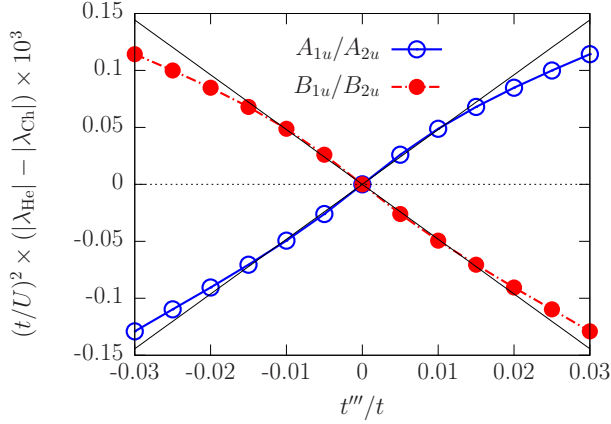}
\caption{Differences between eigenvalues of the effective two-particle interaction in the Cooper pairing channel computed within weak-coupling RG.
Thin black lines are guides for the eye to show the linear behavior at small $t^{\prime\prime\prime}/t$. 
$\lambda_{\mathrm{He}}$ ($\lambda_{\mathrm{Ch}}$) is the eigenvalue for p-wave helical (chiral) pairing states.  
The splitting, $\delta T_c$, of $T_c$ between chiral and helical states is given by $\delta T_c/T_c \propto (|\lambda_{\mathrm{He}}|-|\lambda_{\mathrm{Ch}}|)$, to linear order in $\delta T_c/T_c$. 
The normal state band parameters, other than $t^{\prime\prime\prime}$ and $\eta$, here and elsewhere, are identical to those in Ref.~\onlinecite{Scaffidi2014}.   Here, we choose $\eta = 0.1t$ and $J/U$ = 0.}
\label{fig:split_t}
\end{figure}

Since $\delta \mathcal{F}$ in Eq.~\eqref{eq:deltaFtppp} preserves the pseudospin rotation symmetry in the $xy$-plane, it splits the four p-wave helical states
into two groups, $\{A_{1u},A_{2u}\}$ and $\{B_{1u},B_{2u}\}$. The two states in each group are related to each other by a four-fold pseudospin rotation about $z$.
To leading order in $t^{\prime\prime\prime}/t$, the splitting of $T_c$ between the two groups is $\delta T_c \propto |a_2| \propto |t^{\prime\prime\prime}/t|$.
Since $\delta \mathcal{F}$ does not have any term that splits chiral states from helical states, the transition temperature of the chiral states,
$T_c^{E_u}$, stays half way in between that of the two helical state groups, $T_c^{A_{1u}/A_{2u}}$ and $T_c^{B_{1u}/B_{2u}}$.
We confirm these conclusions with a numerical weak-coupling RG calculation following Refs.~\onlinecite{Raghu2010,Scaffidi2014}.
The results are shown in Fig.~\ref{fig:split_t}. At larger $t^{\prime\prime\prime}/t$, the splitting between helical and chiral states
has deviations from the linear dependence on $t^{\prime\prime\prime}/t$ arising from higher order contributions of $\delta \widetilde{H}_K(\vec{k})$
to $\delta \mathcal{F}$, which lead to terms, $f_{2}^{\mathrm{SOC},4}+f_{2}^{\mathrm{SOC},5}$ and $f_{2}^{\mathrm{SOC},1}$, in $\delta \mathcal{F}$. These terms
leave the degeneracy in each of two helical state groups intact since the pseudospin rotational symmetry around $z$ remains; however, they make the relation $T_c^{E_u}=\big\{ T_c^{A_{1u}/A_{2u}}+T_c^{B_{1u}/B_{2u}} \big\}/2$
only an approximation. Since in $\mathrm{Sr_2RuO_4}$, $|t^{\prime\prime\prime}|/t\sim \eta/t$ is small, we expect
$T_c^{E_u} \approx \big\{ T_c^{A_{1u}/A_{2u}}+T_c^{B_{1u}/B_{2u}} \big\}/2$ to hold, as seen in Fig.~\ref{fig:split_t}.

One conclusion of the above analysis is that the chiral pairing states are never stabilized by the $t^{\prime\prime\prime}$ induced terms.
A similar conclusion was obtained in Ref.~\onlinecite{Annett2006} for a different interaction model within a mean-field analysis.

Since the relative stability between chiral and helical states will be affected by other SOC induced terms, which will be analyzed in detail
in Sec.~\ref{sec:SOCJ}, it is important to understand the SOC dependence of $\delta \mathcal{F}$ in Eq.~\eqref{eq:deltaFtppp}.
Following Ref.~\onlinecite{Annett2006} we go back to the original Hamiltonian before the $\mathcal{U}$ transformation in terms of actual spin.
To linear order in $\eta/t$, the change of the GL free energy due to nonzero SOC is given by $\delta \mathcal{F} = 2 \eta \langle \vec{L}\cdot \vec{S} \rangle$,
where the average $\langle \cdots \rangle$ is evaluated in a mean field pairing state obtained at zero SOC.
From the analysis of $\delta \widetilde{H}_K(\vec{k})$ in pseudospin space, we know that $\delta \mathcal{F}$ is invariant under pseudospin rotations about $z$; it is also invariant under actual spin rotations about $z$ since the pseudospin and actual spin $z$-directions are the same. Therefore, in $\langle \vec{L}\cdot \vec{S}\rangle$, $\langle L^x S^x + L^y S^y\rangle \equiv 0$.
Hence,
\begin{gather}\label{eq:deltaFtppp2}
\delta \mathcal{F} = 2 \eta \langle L^z S^z \rangle = i \eta \bigg\{ n^{\uparrow\uparrow}_{12} - n^{\downarrow\downarrow}_{12} - n^{\uparrow\uparrow}_{21} + n_{21}^{\downarrow\downarrow} \bigg\},
\end{gather}
where $n^{\uparrow\uparrow}_{12}\equiv \sum_i \langle c^\dagger_{i 1 \uparrow} c_{i 2 \uparrow} \rangle$ are the
single-particle density matrices off-diagonal in the orbital index. At zero order in $\eta$, the mean field Hamiltonian for the chiral pairing states
are symmetric with respect to spin $\uparrow \leftrightarrow \downarrow$.  Consequently, the linear in $\eta$ term in $\delta \mathcal{F}$ vanishes and 
$\delta \mathcal{F} \sim \mathcal{O}(\eta^2)$.  On the other hand, for the four helical pairing states, $\delta \mathcal{F} \sim \mathcal{O}(\eta)$
in general, if the superconducting order parameters on the $\alpha$ and $\beta$ bands are not identically
zero when $\eta=0$. This linear dependence has been emphasized in Ref.~\onlinecite{Annett2006}.

We calculate the $\eta$ dependence of $\delta \mathcal{F}$ for our model in weak-coupling RG. In the case of $J=0$, we actually find that $\delta \mathcal{F}
\propto (\eta/t)^2 $ rather than $\propto \eta/t$. This comes from a complete decoupling between the $\alpha+\beta$ and $\gamma$ bands
when $J=J^\prime=\eta=0$, which makes all the density matrices in Eq.~\eqref{eq:deltaFtppp2} identically zero and invalidates the above argument for the linear in $\eta$
dependence (for details see Appendix~\ref{app:etaJeq0}).
When $J\ne 0$, the three bands are coupled and, indeed, we find the leading $\eta/t$ dependence of $\delta \mathcal{F}$ linear, as shown in Fig.~\ref{fig:split_etaJ2}. 

To summarize, the presence of $g(\vec{k})$ and $\eta$ induces a pseudospin $SU(2)$ symmetry breaking term in the GL free energy, given in Eq.~\eqref{eq:deltaFtppp},
which lifts the degeneracy among different p-wave states. This term always favors helical states over the chiral states. It is invariant under the pseudospin
rotations along $z$ that preserves the degeneracy between $A_{1u}$ and $A_{2u}$, and that between $B_{1u}$ and $B_{2u}$. The splitting
between the two helical state groups is $\delta T_c \propto  t^{\prime\prime\prime} \eta /t$, to leading order in $t^{\prime\prime\prime}/t$ and $\eta/t$.
In the special case of $J=J^\prime=0$, the splitting is $\propto t^{\prime\prime\prime} \eta^2/t^2$. Interestingly, the necessary ingredients, $t^{\prime \prime\prime}$ and $\eta$, for the splitting identified here are the same as those responsible for a spin Hall effect discussed in Ref.~\onlinecite{Imai2012}, suggesting that the two may be intimately connected.


\subsection{SOC induced terms due to finite $J=J^\prime$ but with $g(\vec{k})\equiv 0$} \label{sec:SOCJ}
In this section, we analyze the pseudospin rotational breaking terms due to finite $J$ and $\eta$, while keeping $g(\vec{k})\equiv 0$.

\subsubsection{Pseudospin $SU(2)$ breaking terms}
When $J\ne 0$, applying the $\mathcal{U}$ transformation in Eq.~\eqref{eq:U} to $\widetilde{V}$ in Eq.~\eqref{eq:Kanamori3} changes the form of $\widetilde{V}$ and leads to
\begin{gather} \label{eq:Kanamori4}
\widetilde{\widetilde{V}} \equiv \mathcal{U}^\dagger \widetilde{V} \mathcal{U} = \widetilde{\widetilde{V}}_{\mathrm{inv}}
+\widetilde{\widetilde{V}}_J + \widetilde{\widetilde{V}}_{J^\prime},
\end{gather}
where $ \widetilde{\widetilde{V}}_{\mathrm{inv}} = \mathcal{U}^\dagger \; ( \widetilde{V}_U+\widetilde{V}_{U^\prime}) \; \mathcal{U}=  \widetilde{V}_U+\widetilde{V}_{U^\prime}$ is still pseudospin $SU(2)$ invariant. The other two terms are
\begin{widetext}
\begin{subequations}
\begin{align}
\widetilde{\widetilde{V}}_J & = \mathcal{U}^\dagger \widetilde{V}_J  \mathcal{U} = -J \sum_{i} \bigg\{
 \bigg[
 S^x_{i2} S^x_{i3} +  S^{y}_{i1}  S^y_{i3}  + S^z_{i1} S^z_{i2}
 \bigg]
 -
 \bigg[
 S_{i1}^x S^x_{i2}  + S^x_{i1} S^x_{i3} + S^y_{i1} S^y_{i2} + S^y_{i2} S^y_{i3}
 + S^z_{i1}  S^z_{i3} + S^z_{i2} S^z_{i3}
 \bigg]
 \bigg\},  \label{eq:VJ} \\
 \widetilde{\widetilde{V}}_{J^\prime} & =  \mathcal{U}^\dagger \widetilde{V}_{J^\prime} \mathcal{U} =
\frac{J^\prime}{2} \sum_{i, \sigma \ne \sigma^\prime}\bigg\{
- \sum_{ a\ne b=\{1,2\}} -  \sum_{ a\ne b=\{2,3\}} + \sum_{a\ne b=\{1,3\}} \bigg\} \quad
a^\dagger_{i a  \sigma} a^\dagger_{i a \sigma^\prime}  a_{i b \sigma^\prime}  a_{i b \sigma}.
\end{align}
\end{subequations}
\end{widetext}

The $\mathcal{U}$ transformation shifts the SOC induced effect of spin rotational symmetry breaking from the kinetic energy part of the Hamiltonian
to the interaction part. Note that the kinetic energy part becomes pseudospin $SU(2)$ invariant after the transformation.
Since each term of the original interaction $V$ in Eq.~\eqref{eq:Kanamori2} is $SU(2)$ spin rotational invariant, we can identify
the pseudospin $SU(2)$ rotational symmetry breaking terms in  $\widetilde{\widetilde{V}}$ as
\begin{widetext}
\begin{align} \label{eq:JSOC}
\delta \widetilde{\widetilde{V}} & = -2 J \sum_{i} 
 \bigg[
 S^x_{i2} S^x_{i3} +  S^{y}_{i1}  S^y_{i3}  + S^z_{i1} S^z_{i2}
 \bigg]  + J^\prime \sum_{i, \sigma \ne \sigma^\prime}  \sum_{ a\ne b=\{1,3\}} a^\dagger_{i a  \sigma} a^\dagger_{i a \sigma^\prime}  a_{i b \sigma^\prime}  a_{i b \sigma}.
\end{align}
\end{widetext}
In this equation the $J^\prime$ term alone does not lift the degeneracy
among different p-wave pairing states. This can be proved within weak-coupling RG and RPA approximations by examining diagramatic
contributions to helical and chiral states at each order in interaction. There is a one-to-one correspondence between the two
contributions that contain $J^\prime$, if $J=0$.
This result is consistent with Ref.~\onlinecite{Yanase2003}, where a direct perturbation, up to second order in both interaction and SOC, shows that the SOC induced terms
to the effective interaction in the Cooper pairing channel necessarily depend on $J$ when $g(\vec{k})\equiv 0$. 
We have also verified the above conclusion in our numerical weak-coupling RG and RPA calculations. Therefore, within linear order in $J$ ($=J^\prime$),
we can drop the $J^\prime$ term in Eq.~\eqref{eq:JSOC}.  

\subsubsection{GL free energy terms due to  $\delta \widetilde{\widetilde{V}}$}

$\delta \widetilde{\widetilde{V}}$ in Eq.~\eqref{eq:JSOC} does not completely break the pseudospin $SU(2)$ rotational symmetry.
Mirror reflections about the $xz$- and $yz$-planes, denoted as $\mathcal{M}_{xz}^{\hat{S}}$ and $\mathcal{M}_{yz}^{\hat{S}}$ respectivly, leave
$\delta \widetilde{\widetilde{V}}$ invariant. This holds even if the $J^\prime$ term in Eq.~\eqref{eq:JSOC} is taken into account.
$\mathcal{M}_{xz}^{\hat{S}}$ and $\mathcal{M}_{yz}^{\hat{S}}$ are therefore symmetries of the whole microscopic Hamiltonian. In Table~\ref{tab:SOCcomplete},
the only terms compatible with these symmetries are $f_2^{\mathrm{SOC},1},f_2^{\mathrm{SOC},4}$, and $f_2^{\mathrm{SOC},5}$. Therefore, in general, 
the GL free energy due to $\delta \widetilde{\widetilde{V}}$ is given by
\begin{align} \label{eq:deltaFJ}
\delta \mathcal{F} & =  a_1 \; f_2^{\mathrm{SOC},1}  + a_4 \; f_2^{\mathrm{SOC},4} + a_5 \; f_2^{\mathrm{SOC},5} \nonumber \\   
 & = \frac{2a_1 +a_4 +a_5}{4} \big[ f_2^{\mathrm{SOC},1}  + f_2^{\mathrm{SOC},4} + f_2^{\mathrm{SOC},5}\big] \nonumber \\
  & + \frac{2a_1 -a_4 -a_5}{4} \big( f_2^{\mathrm{SOC},1}  - f_2^{\mathrm{SOC},4} - f_2^{\mathrm{SOC},5}\big) \nonumber \\
  & + \frac{a_4 -a_5}{2} \big( f_2^{\mathrm{SOC},4} - f_2^{\mathrm{SOC},5}\big).
\end{align}
where $\{a_1, a_4, a_5\}$ are three coefficients that shift the $T_c$ away from $T_c^0$.
 In $\delta \mathcal{F}$, $( f_2^{\mathrm{SOC},1}  + f_2^{\mathrm{SOC},4} + f_2^{\mathrm{SOC},5})$ is trivial and  shifts the $T_c$ of all p-wave pairing states equally.
 To leading order in $J/U$, $\{a_1,a_4,a_5\} \propto J$.
 \footnote{{
 The leading order contribution to $\delta \mathcal{F}$ in a weak-coupling theory comes from a second order perturbation result, 
$\delta \mathcal{F}= \langle \widetilde{\widetilde{V}}_{\mathrm{inv}} \; \widetilde{G}_4 \;\delta \widetilde{\widetilde{V}} \rangle$,
which is second order in $\widetilde{\widetilde{V}}$ but first order in $J/U$.  Here $\langle \cdots \rangle$ means being averaged in a mean-field p-wave pairing state
obtained at $J=0$, and $\widetilde{G}_4$ is the four-point Green's function defined for the normal state Hamiltonian after the  $\mathcal{U}$ transformation,
$\widetilde{H}_K$ given in Eq.~\eqref{eq:HKtilde}. 
Note that the first order perturbation contribution, $\langle \delta \widetilde{\widetilde{V}}\rangle$, is identically zero for a $p$-wave pairing state because the interaction $\delta \widetilde{\widetilde{V}}$ is purely on-site. 
However, because the pseudospin rotational symmetry property of $\delta \mathcal{F}$
is completely dictated by $\delta \widetilde{\widetilde{V}}$, in the main text we simply focused on $\delta \widetilde{\widetilde{V}}$, rather than the more complicated $\widetilde{\widetilde{V}}_{\mathrm{inv}} \; \widetilde{G}_4 \;\delta \widetilde{\widetilde{V}}$
}}
 $( f_2^{\mathrm{SOC},1}  - f_2^{\mathrm{SOC},4} - f_2^{\mathrm{SOC},5}) $ splits the chiral state away from helical ones, 
while $(f_2^{\mathrm{SOC},4} - f_2^{\mathrm{SOC},5})$ breaks the degeneracy among the four helical p-wave states, splitting them into two groups,
$\{A_{1u},B_{1u}\}$ and $\{A_{2u},B_{2u}\}$. Within each group the two states are connected by $\mathcal{M}_{xz}^{\hat{S}}$ and $\mathcal{M}_{yz}^{\hat{S}}$, 
and therefore remain degenerate. In terms of the components of the order parameter matrix $\hat{\Delta}$, 
$f_2^{\mathrm{SOC},4} - f_2^{\mathrm{SOC},5}=(-1/2)\big\{ [\Delta_{\uparrow\uparrow,x}^* \Delta_{\downarrow\downarrow,x} -\Delta_{\uparrow\uparrow,y}^*
\Delta_{\downarrow\downarrow,y}] + c.c. \big\}$. This term was identified in Ref.~\onlinecite{Yanase2003,Yanase2014} 
using a direct expansion in the SOC, while our analyses here are based on symmetries of the model.

Again, it is important to understand the SOC dependence of $\delta \mathcal{F}$ in Eq.~\eqref{eq:deltaFJ}. For that we go back to the original Hamiltonian
written in terms of the actual spin. 
As mentioned previously, the linear order in $\eta/t$ contribution to the GL free energy comes from
$\delta \mathcal{F} = 2 \eta \langle \vec{L}\cdot \vec{S} \rangle$, where $\vec{S}$ is the actual spin operator, not pseudospin. 
However, $\langle \vec{L}\cdot \vec{S} \rangle \equiv 0$ because of the three remaining
mirror reflection symmetries in the pseudospin space, $\{\mathcal{M}_{xz}^{\hat{S}},\mathcal{M}_{yz}^{\hat{S}},\mathcal{M}_{xy}^{\hat{S}}\}$, 
which imply the same symmetries for the actual spin, since the $\{x,y,z\}$-directions are identical in the pseudospin and actual spin spaces. On the other hand, these symmetries do not prohibit a second order in $\eta/t$ term, 
$\delta \mathcal{F} \propto \langle (2 \eta \; \vec{L}\cdot \vec{S})^2 \cdots \rangle$, where $\cdots$ here stands for $\eta$ independent operators that have a dimension of energy inverse. Therefore, in Eq.~\eqref{eq:deltaFJ}, 
the GL expansion coefficients $\{a_1,a_4,a_5\} \propto (\eta/t)^2$ to leading order in $\eta/t$. 

\subsubsection{Numerical results}
\begin{figure}[tp]
\centering
\includegraphics[width=\linewidth,trim={22mm 0mm 0mm 0mm},clip]{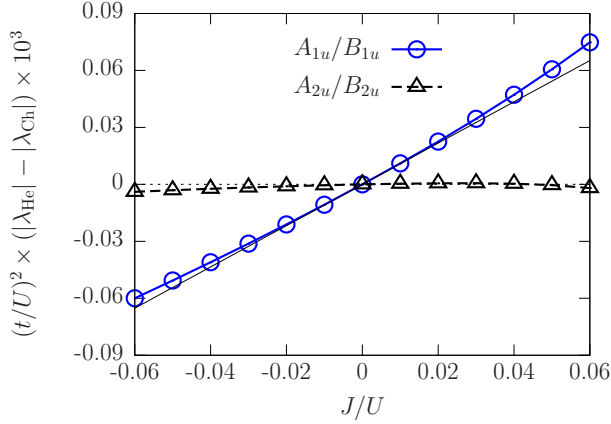}
\caption{$J/U$ dependence of the splitting between helical and chiral p-wave pairing states in weak-coupling RG. $t^{\prime\prime\prime}=0$ and $\eta=0.1t$.
The splitting is linear in $J/U$ at small $J/U$. Note that the two helical states, $\{A_{2u},B_{2u}\}$, are almost degenerate with the chiral state, which is accidental
and not robust to band parameter changes.  }
\label{fig:JHSplit0}
\end{figure}

We confirm the above conclusions with weak-coupling RG calculations, where the details of the calculation follow Refs.~\onlinecite{Raghu2010,Scaffidi2014}.
Fig.~\ref{fig:JHSplit0} shows the numerical results of the splitting between helical and chiral states as a function of $J/U$ for fixed $\eta/t=0.1$. 
At $J/U=0$, all p-wave pairing states are degenerate, even though $\eta \ne 0$, consistent with the conclusion
obtained in Sec.~\ref{sec:degeneracy}. At finite $J/U$, the degeneracy between chiral and helical states is lifted. The four
helical states are split into two groups of two degenerate states.  The splitting of $T_c$ between the chiral states and the $\{A_{1u}, B_{1u}\}$ 
group is indeed $\propto J/U$ to leading order, as predicted. Interestingly, the other group, $\{A_{2u},B_{2u}\}$, remains almost degenerate with the chiral states
even at finite $J/U$, which is, however, not robust to changes of normal state band dispersions. 

\begin{figure}[tp]
\centering
\includegraphics[width=\linewidth,trim={22mm 0mm 0mm 0mm},clip]{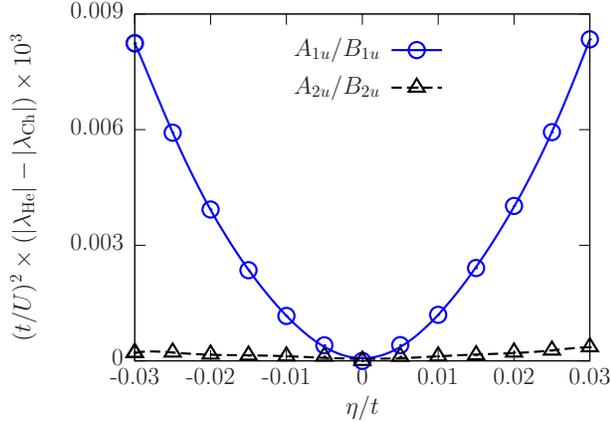}
\caption{$\eta/t$ dependence of the spliting between helical and chiral states in weak-coupling RG. $t^{\prime\prime\prime}=0$ and $J/U=0.06$.
The splitting is $\propto (\eta/t)^{2}$ to leading order in $\eta/t$. Again, due to the near-degeneracy between the $\{A_{2u},B_{2u}\}$ and chiral states for these band parameters,
the quadratic dependence of the splitting in $\eta/t$ is difficult to discern. }
\label{fig:split_etaJ3}
\end{figure}

Fig.~\ref{fig:split_etaJ3} shows our weak-coupling RG results for the SOC dependence of $|\lambda_{\mathrm{He}}|-|\lambda_{\mathrm{Ch}}|$. Within numerical errors,
$|\lambda_{\mathrm{He}}|-|\lambda_{\mathrm{Ch}}| \propto (\eta/t)^2$, in agreement with the above analytical analysis. 

A summary of the main results obtained in this section is: finite $J=J^\prime$ and $\eta$ induce pseudospin rotational breaking terms in the GL free energy as given in Eq.~\eqref{eq:deltaFJ},
which lift the degeneracy among different p-wave pairing states. The splitting of $T_c$ between different p-wave states is $\delta T_c \propto
(J/U) \, \eta^2/t$, to leading order in $J/U$ and $\eta/t$. The degeneracy between $A_{1u}$ and $B_{1u}$, and that between $A_{2u}$ and $B_{2u}$,
remains due to pseudospin mirror reflection symmetries in the $xz$ and $yz$ planes. The terms in Eq.~\eqref{eq:deltaFJ} can favor either chiral or helical states,
depending on the magnitudes of the two coefficients, $(2a_1 -a_4 -a_5)/4$ and $(a_4-a_5)/2$, which in turn depend on the normal
state band structures. If $a_1<\min\{a_4,a_5\}$, then $(2a_1 -a_4 -a_5)/4 < -|a_4-a_5|/4$ and the chiral states are stabilized.

\subsection{Results for both $g(\vec{k})\ne 0$ and $J\ne 0$}
\begin{figure}[tp]
  \centering
    \includegraphics[width=\linewidth,trim={22mm 0mm 0mm 0mm},clip]{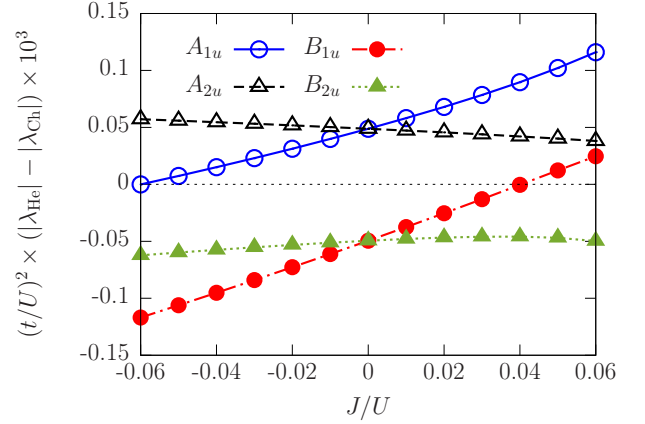}
  \caption{$J/U$ dependence of the splitting between different helical and chiral p-wave states.  $t^{\prime\prime\prime}=0.01t$ and $\eta=0.1t$. 
  The splitting at small $J/U$ is dominated by the linear in $t^{\prime\prime\prime}$ effect discussed in Sec.~\ref{sec:SOCtppp}, which always stabilizes helical states.}
  \label{fig:JHSplitting}
\end{figure}

When both $g(\vec{k})$ and $J=J^\prime$ are non-zero, the SOC induced GL free energy is given by the sum of Eqs.~\eqref{eq:deltaFtppp}
and~\eqref{eq:deltaFJ}. However, the GL free energy expansion coefficients for each $f_{2}^{\mathrm{SOC},j}$ are different from those in  Eqs.~\eqref{eq:deltaFtppp}
and~\eqref{eq:deltaFJ} because of additional contributions that depend on both $t^{\prime\prime\prime}$ and $J$.
The degeneracy among all p-wave pairing states is lifted
except the one between the two chiral states with opposite chirality within the $E_{u}$ representation, as seen in Fig.~\ref{fig:JHSplitting}. 
Because of the near-degeneracy seen in Fig.~\ref{fig:JHSplit0}, the splitting between $A_{2u}$ (or $B_{2u}$) and chiral states is dominated
by the $t^{\prime\prime\prime}$ term at small $J/U$. An implication is that, with both $J$ and $t^{\prime\prime\prime}$ present, the dominant
p-wave pairing state in the small $J/U$ and $t^{\prime\prime\prime}/t$ parameter space regime will be always helical, rather than chiral, regardless of whether 
the splitting, $|\lambda_{\mathrm{He}}|-|\lambda_{\mathrm{Ch}}|$, for the other two helical states, $\{A_{1u},B_{1u}\}$, is $\propto A J/U$ with a positive slope $A>0$, as 
seen in Fig.~\ref{fig:JHSplitting}, or with $A<0$. When $t^{\prime\prime\prime}/t$ becomes larger, the splitting between
$\{A_{2u},B_{2u}\}$ and chiral states can pick up a significant $J/U$ linear dependence because of cross dependent terms. 

\begin{figure}[tp]
\centering
\includegraphics[width=\linewidth,trim={22mm 0mm 0mm 0mm},clip]{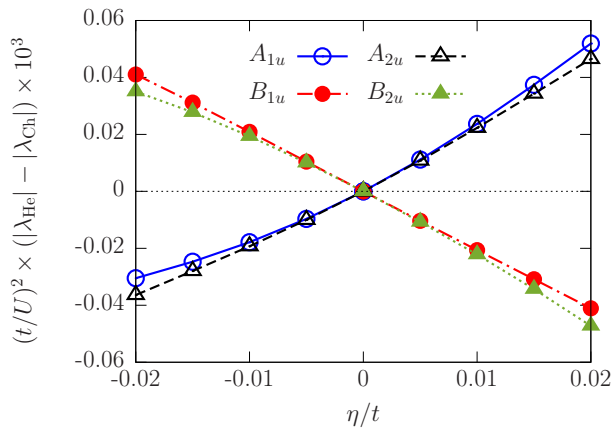}
\caption{$\eta/t$ dependence of the splitting between helical and chiral p-wave states in weak-coupling RG.  $t^{\prime\prime\prime}=0.1t$ and $J/U=0.11$. The leading SOC dependence
is linear at small $\eta/t$. }
\label{fig:split_etaJ2}
\end{figure}

Some of the conclusions derived in Sec.~\ref{sec:SOCtppp} and ~\ref{sec:SOCJ} still hold when both $t^{\prime\prime\prime}$ and $J$ are present.
For example, the leading SOC dependence of the splitting between different p-wave pairing states is linear due to the $g(\vec{k})$ induced terms, 
as shown in Fig.~\ref{fig:split_etaJ2}. These terms are $\propto t^{\prime\prime\prime} \eta/ t^2$ to leading order in $t^{\prime\prime\prime}/t$.

\section{Stability of chiral p-wave pairing} \label{sec:SRO}
\begin{figure}[tp]
\centering
\includegraphics[width=\linewidth]{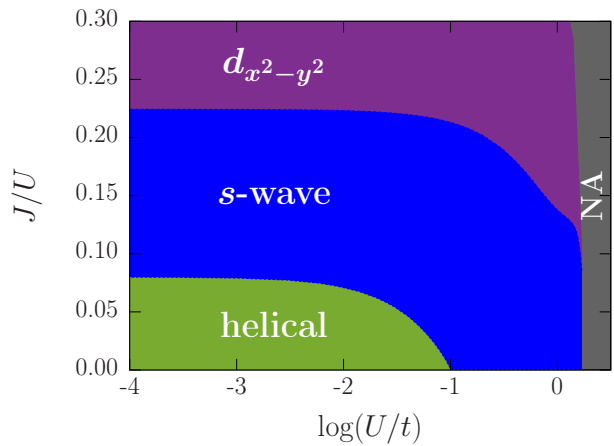}
\caption{Phase diagram obtained within RPA for different $J/U$ and $U/t$. The RPA breaks down in the ``NA" regime. The three intra-orbital hoppings in Eq.~\eqref{eq:HK} are
$\epsilon_{xz (yz)}(\vec{k}) = -2t \cos{k_{x(y)}} -2t^{\perp}\cos{k_{y(x)}}-\mu$,
$\epsilon_{xy}(\vec{k}) = -2t'(\cos{k_x}+\cos{k_x})-4t''\cos{k_x}\cos{k_y} -2t''''(\cos{2k_x}+\cos{2k_x})-\mu$. 
We choose the band parameters,  $(t,~t^{\perp},~t',~t'',~t''',~t'''',~\mu,~\eta) = (1,~0.1,~0.8,~0.3,~0.05,~-0.015,~1.075,~0.2) $, such that
the resulting Fermi surfaces fit recent ARPES data~\cite{Tamai2019}.}
\label{fig:phasediagram}
\end{figure}

The analysis of Sec.~\ref{sec:microscopic} shows that, within the current 2D three-band model with an on-site Kanamori-Hubbard interaction, 
the dominant pairing is always helical, rather than chiral, at small $J/U$ and $U/t$ where p-wave pairing is favored within the weak-coupling approximation~\cite{Scaffidi2014,Zhang2018,Roising2019}. 
On the other hand, at large $J/U$, pseudospin singlet pairing takes over~\cite{Scaffidi2014,Zhang2018,Roising2019}. 
Therefore we expect the phase diagram, in the parameter space spanned by $J/U$ and $U/t$, to be dominated by helical p-wave and singlet pairing states for physical band parameters describing Sr$_2$RuO$_4$, 
where inter-orbital hybridization between $d_{xz}$ and $d_{yz}$ orbitals can not be neglected. 
This expectation is confirmed by our RPA calculations, which give the phase diagram shown in Fig.~\ref{fig:phasediagram}. Details of the RPA calculation follow those found in Refs.~\onlinecite{Zhang2018,Romer2019}. 
The RPA breaks down for $U/t\gtrsim \mathcal{O}(1)$
due to an instability inherent in this approximation, 
~\footnote{
The eigenvalue of the effective interaction in the Cooper pairing channel, $\lambda$, diverges where RPA breaks down and for Fig.~\ref{fig:phasediagram}, we cut off the phase diagram at $(t/U)^2|\lambda|=0.2$. The boundary is insensitive to the choice of cut-off (provided it is $\gtrsim \mathcal{O}(1)$), since $\lambda$ diverges rapidly in this region
}
but can give reliable results even beyond the weak-coupling regime, $U/t\ll 1$.~\cite{Romer2019a}
In Fig.~\ref{fig:phasediagram}, there is no trace of chiral pairing even at an intermediate value of $J/U$. 
In this phase diagram, the helical state order parameter realized is $\vec{d}(\vec{k})=\hat{x} k_x + \hat{y} k_y$ ($A_{1u}$), and the $s$ and $d_{x^2-y^2}$ wave order parameters
belong to the irreducible representation $A_{1g}$  and  $B_{1g}$, respectively, of the $D_{4h}$ group. However, they are not simple lowest harmonic functions, but are highly anisotropic, similar to those found in Refs.~\onlinecite{Scaffidi2014, Zhang2018}. 
In each phase of the phase diagram, the ratio of the gap magnitude on different bands depends on both $J/U$ and $U/t$. 
However, unlike Ref.~\onlinecite{Scaffidi2014} where the $\alpha$ + $\beta$ always dominate  when the favored pairing symmetry is helical, 
we find that the dominant band in the helical phase is $\gamma$ when both $J/U$ and $U/t$ are small, while it changes to $\alpha+\beta$ at larger $J/U$ or $U/t$. 

Since chiral p-wave pairing states have been previously found in various numerical calculations using the same model~\cite{Yanase2003,
Scaffidi2014,Zhang2018,Romer2019}, we comment on these. In Ref.~\onlinecite{Yanase2003}, the dominant pairing instability was calculated by solving Eliashberg
equations with an effective pairing interaction derived from a perturbation theory up to second order in the bare interaction. Chiral p-wave was found to be
the dominant channel when the Eliashberg equations were solved only for the $\gamma$ band, while the coupling between $\gamma$ and $\alpha+\beta$ bands due to the effective
interaction was neglected. However, this coupling can have significant effects on the ratio between the gap magnitudes of the two sets of bands~\cite{Scaffidi2014,Zhang2018},
which in turn can impact the relative stability between helical and chiral p-wave pairing states. This can explain the difference between our numerical results and those
in Ref.~\onlinecite{Yanase2003}. 
Ref.~\onlinecite{Scaffidi2014} is a weak-coupling RG calculation, where chiral p-wave states have been found 
near $J/U=0$ with a nonzero $t^{\prime\prime\prime}/t=0.01$. However, this is inconsistent with our analytical analyses of the $t^{\prime\prime\prime}$ effect in Sec.~\ref{sec:SOCtppp}
and also inconsistent with our numerical results in Fig.~\ref{fig:phasediagram}. Ref.~\onlinecite{Zhang2018} is an RPA calculation based on the same model. The phase diagrams obtained in the weak-coupling limit
are similar to those in Ref.~\onlinecite{Scaffidi2014}. In particular, there is a significant portion of the phase diagram at small $J/U$ and
$U/t$, where chiral p-wave pairing dominates. However, we note that Eq. (S13) of Ref.~\onlinecite{Zhang2018} takes the real part of the effective interaction. 
In the presence of SOC, this suppresses the $t^{\prime\prime\prime}$ induced terms that we have identified in Eq.~\eqref{eq:deltaFtppp}, 
which favors helical over chiral states. This may explain the discrepancy between our RPA phase diagram in Fig.~\ref{fig:phasediagram} and those 
in Ref.~\onlinecite{Zhang2018}. In Ref.~\onlinecite{Romer2019}, a similar RPA calculation was performed at relatively large $U$ for different Fermi surface geometries.
Chiral p-wave pairing has been found only at large SOC for the Fermi surface geometry where the $\gamma$ band touches the zone boundary. However, in that calculation,
the inter-orbital hybridization $t^{\prime\prime\prime}$ was set to zero, which completely leaves out the terms in Eq.~\eqref{eq:deltaFtppp}. 
Physically we do not expect this hybridization to be vanishingly small, given that it is between orbitals on two next-nearest neighboring sites.
Including a small $t^{\prime\prime\prime}=0.01 t$ suppresses the chiral p-wave pairing, giving way to helical states. We have verified this with RPA calculations in a parameter regime that overlaps with those of  Ref.~\onlinecite{Romer2019} 
and found results that are consistent with our analytical analysis.  Furthermore, we find the stability of chiral p-wave in this parameter regime requires fine-tuning, in that a small change in parameters renders this phase unstable.

Given the difficulty of stabilizing a chiral p-wave state within the current model, we wonder what ingredients can favor a chiral p-wave state if we go beyond this model. 
There are at least two possibilities to consider: (1) three-dimensional effects on the normal state Fermi surface; (2) longer range 
off-site interactions.

In a 3D model with the same on-site Kanamori-Hubbard interaction, like the one used in Ref.~\onlinecite{Roising2019},
two additional inter-orbital hybridization terms appear in the normal state Hamiltonian, $t_{xz,xy}$ ($t_{yz,xy}$) 
between $d_{xz}$ ($d_{yz}$) and $d_{xy}$ orbitals, in addition to the $t^{\prime\prime\prime}$ that we have already considered. 
The two inter-layer hybridizations $t_{xz,xy}$ and $t_{yz,xy}$, combined with the finite SOC, can mix an out-of-plane component $k_z (\hat{x},\hat{y})$ in the $\vec{d}(\vec{k})$ vector of the chiral p-wave pairing state within the $E_u$ representation~\cite{Sigrist1991, Huang2018, Pustogow2019}, which shifts the $T_c$ of the chiral p-wave state. However, this mixing is small because of its dependence on small parameters $\eta/t$ and $t_z/t$, where $t_z\equiv \max\{ |t_{yz,xy}|, |t_{xz,xy}|\}$. The mixing vanishes if either $\eta=0$ or $t_z=0$ so that, to leading order
in $\eta/t$ and $t_z/t$, it is $\propto \eta t_z/t^2$. The resulted critical temperature shift from the mixing can be estimated by a second order non-degenerate perturbation theory and the result is $\delta T_c/T_c \propto (\eta t_z/t^2)^2$. Detailed discussions can be found in Appendix~\ref{app:SOC3D}. This shift is negligible compared to the effects of other SOC induced terms on $T_c$ that we discussed in Sec.~\ref{sec:microscopic}. Therefore, we ignore the possible mixing in the following. 

Then we can easily generalize our 2D analyses to the 3D
model. If we set all inter-orbital hybridizations and $J$ to zero, the same derivations as in Sec.~\ref{sec:degeneracy} lead to the same conclusion that
all p-wave pairing states remain degenerate even with $\eta \ne 0$. 
Note that in this case there is no mixing between the in-plane and out-of-plane pairings because the full pseudospin $SU(2)$ symmetry is still preserved. 
SOC induced terms by each inter-orbital hybridization can be analyzed similarly
following Sec.~\ref{sec:SOCtppp}. The two additional hybridizations, $t_{xz,xy}$ and $t_{yz,xy}$, add two additional terms to the $\delta \widetilde{H}_K (\vec{k}) $
in Eq.~\eqref{eq:deltaHK} that are $\propto t_{xz,xy} \sin k_z/2 \cos k_x/2 \sin k_y/2 \cdots $ and $\propto t_{yz,xy} \sin k_z/2 \cos k_y/2 \sin k_x/2 \cdots$,
respectively~\cite{Roising2019}. However, the leading order GL free energy from these two terms vanish in $\delta \mathcal{F} =\langle \delta \widetilde{H}_K (\vec{k}) \rangle$ after the
$\vec{k}$ average, since we have ignored a possible mixing of the out-of-plane $p_z$ pairing component, and the odd $k_z$ dependence of those two terms can not be compensated by any other term in the mean field Hamiltonian of a $p_x$ or $p_y$ pairing state.
Therefore, to linear order in $t_{xz,xy}/t$ or $t_{yz,xy}/t$, which are expected to be even smaller than $t^{\prime\prime\prime}/t$, we can drop those additional hybridizations
in the normal state Hamiltonian.  Then the analyses of the $t^{\prime\prime\prime}$ and $J$ induced terms are identical to those in Sec.~\ref{sec:SOCtppp}
and ~\ref{sec:SOCJ}. Therefore, the conclusions obtained in the 2D analysis can be directly applied to the 3D model. In other words, the three-dimensional
effect of the FS does not help stabilize a chiral p-wave pairing state, consistent with the 3D weak-coupling RG numerical results obtained in
Ref.~\onlinecite{Roising2019}, where helical states have been found to dominate over chiral p-wave pairing at $J/U$ all the way up to $J/U=0.2$. 

Another possibility is to consider longer-range off-site interaction models~\cite{Koikegami2003,Ng2000b}. 
Ref.~\onlinecite{Ng2000b} considered such a model with attractive nearest neighbor interactions, and, indeed, chiral p-wave pairing states were found to be stabilized in some regime of the pairing interaction parameter space. 
However, the solutions were obtained under the assumption that the p-wave pseudo-spin triplet channel is favored over singlet channels. 
In Ref.~\onlinecite{Koikegami2003}, the authors studied a nearest neighbor version of the Kanamori-Hubbard interaction model, and found that p-wave pseudospin triplet states are more stable than singlet channels for certain choice of interaction parameters; on the other hand, 
the relative stability among different p-wave pairing states has been completely ignored by simply assuming that chiral p-wave pairing states are favored over helical states. 
In both cases, further investigations beyond the assumptions made here would be needed to establish the stability of chiral p-wave states.

\section{Conclusions} \label{sec:conclusion}
We have conducted a thorough study of the  effect of SOC on the relative stability of different p-wave pairing states for a widely used
microscopic model for $\mathrm{Sr_2RuO_4}$. Our analysis combines a general GL free energy expansion with an analytical study of the symmetry of the microscopic model Hamiltonian.  We give the most general form of the SOC induced quadratic
GL terms that break the pseudo-spin $SU(2)$ rotation symmetry,  identify the relevant GL terms for the microscopic model, and examine their effects on lifting the degeneracy among different p-wave pairing states. The analytical results
are further supported by  our weak-coupling RG and RPA numerical calculations. 

A theme that emerges from this study is that the breaking of $SU(2)$ rotation symmetry in pseudospin space can be quite different from that in the actual spin space; this was also pointed out in Ref.~\onlinecite{Yanase2003}. The former depends on not only the presence of SOC but also other ingredients
of the microscopic Hamiltonian, which in the current model are the inter-orbital hybridization $t^{\prime\prime\prime}$, Hund's coupling $J$, and/or pair hopping. 
The additional dependence on $t^{\prime\prime\prime}$ and $J$ significantly reduces the splitting among different p-wave states for $\mathrm{Sr_2RuO_4}$ since both
$t^{\prime\prime\prime}/t$ and $J/U$ are small.  In the parameter space regime relevant to
$\mathrm{Sr_2RuO_4}$, with finite but small $t^{\prime\prime\prime}/t$ and small $J/U$, we find that the finite $t^{\prime\prime\prime}$ effect tends to dominate and always stabilizes helical states over the chiral ones. We have also generalized our analysis to a 3D model and shown that the existence of inter-orbital hybridizations, in addition to
the $t^{\prime\prime\prime}$ that already exists in 2D models, does not help stabilize the chiral p-wave pairing states, in agreement with the recent numerical study~\cite{Roising2019}. On the other hand, including longer-range interactions may or may not make the chiral states more favorable and requires further investigation. 

Our analysis has resolved some conflicts among different results on the relative stability between helical and chiral p-wave pairing states in the literature.  Since the analysis is largely based on the symmetries of the model and independent of how the model is treated, it also serves as a guide for future studies, both analytical and numerical. Furthermore, the analysis presented here can be adapted to study the effect of SOC on other multi-orbital pseudospin triplet superconductors.  

An outstanding issue in $\mathrm{Sr_2RuO_4}$  is to reconcile theory with the observations of broken time-reversal symmetry~\cite{Luke1998, Xia2006} and a jump in the shear modulus $c_{66}$~\cite{Lupien2002, *Ghosh2019}. Although a chiral p-wave state can explain both, 
here we discuss possible alternative explanations with helical states. 

Given the small splitting among different helical states found here and in previous works~\cite{Scaffidi2014, Zhang2018, Roising2019, Ramires2019},
a possibility to consider is a pair of accidentally or nearly degenerate helical states.
If the two states are close enough to degeneracy, such a pair can lead to either coexistence of different helical state domains~\cite{Khodas2012, Pustogow2019} or a homogeneous time reversal breaking state~\cite{Agterberg1999, Stanev2010,Khodas2012}, depending on microscopic interactions.   A previous analysis of quartic GL terms~\cite{Huang2016} suggests that a homogeneous time reversal symmetry breaking state is almost impossible unless the system is very near or right at the degeneracy point. Moreover, except right at the degeneracy point, this scenario requires two phase transitions with
different $T_c$, which is not observed experimentally. Nevertheless, if two almost degenerate helical orders do form a homogeneous state, this can lead to a jump in $c_{66}$ if the two mixed representations are $ \{ A_{1u}, B_{2u}\}$ or $ \{ A_{2u}, B_{1u}\}$~\cite{Ramires2019}. 
However, within the models studied in this paper, our analysis in Sec.~\ref{sec:microscopic} suggests that residual symmetries in pseudospin space do not naturally lead to a degeneracy between $A_{1u}$ ($A_{2u}$) and $B_{2u}$ ($B_{1u}$).

In the case of coexistence of domains, time reversal breaking is possible at domain walls where two different helical order parameters coexist. However,
since the order parameter mixing is local, the resulted coupling to external probes, such as light in a Polar Kerr measurement~\cite{Xia2006} and the shear strain $\epsilon_{xy}$ in an ultrasound measurement~\cite{Lupien2002}, is also local, which makes it unlikely to be able to account for the experiments.
So a theoretical explanation of both broken time-reversal symmetry and a jump in $c_{66}$  is highly constrained. Further investigations, both experimentally and theoretically, are needed to better assess the possibility of  reconciling the experiments with helical ordered states.

\textit{Note added.} Recently, Ref. \onlinecite{Romer2019} was updated with additional calculations on the effects of the inter-orbital hybridization $t^{\prime\prime\prime}$. Their new results are consistent with our analysis.

\section{Acknowledgments}
We would like to thank Sung-Sik Lee, Wen Huang and Thomas Scaffidi for useful discussions. This research is supported by the National Science and Engineering
Research Council of Canada (NSERC) and the Canadian Institute for Advanced Research (CIFAR).
This work was made possible by the facilities of the Shared Hierarchical Academic Research Computing Network (SHARCNET:www.sharcnet.ca) and Compute/Calcul Canada.

\appendix \label{sec:appendix}

\begin{appendix}

\section{SOC induced GL free energy terms at quadratic order for 2D models} \label{app:SOCDerivation}
As mentioned in the main text, the remaining symmetry group in the presence of SOC is $D_{4h}^{\hat{L}+\hat{S}} \otimes U(1)^C$
for a 2D model.
To derive all possible GL free energy terms at quadratic order for the pseudospin triplet pairing states, we contract
the rank-4 tensor, $(d_i^\mu)^* d_j^\nu$, to a scalar such that it is invariant under all symmetry operations of the above group.
For 2D models, the $xy$-plane mirror reflection symmetry of $D_{4h}^{\hat{L}+\hat{S}}$, denoted as $\mathcal{M}_{xy}^{\hat{S}}$, is operative only on the pseudospin
since there is no $k_z$. $\mathcal{M}_{xy}^{\hat{S}}$ requires that, in $(d_i^\mu)^* d_j^\nu$, either $\{\mu,\nu\}=\{x,y\}$ or $\mu=\nu=z$.

For the case of $\{\mu,\nu\}=\{x,y\}$, there are only four possible independent contractions, given in Table~\ref{tab:contraction}.
\begin{table}[htp]
  \centering
    \caption{All possible contractions of $(d_i^\mu)^* d_j^\nu$ that are invariant under the $D_{4h}^{\hat{L}+\hat{S}} \otimes U(1)^C$ group for $\{\mu,\nu\}=\{x,y\}$.
    For 2D models, $\{i,j\}=\{x,y\}$. }
  \begin{adjustbox}{width=\linewidth}
      \begin{tabular}{@{\hskip 0.05in} c @{\hskip 0.2in} c  @{\hskip 0.05in}}
  \hline
  \hline
  Different contractions & Results \\
  \hline
 $\sum_{\mu \nu i j=\{x,y\}} \; \delta_{\mu i} \delta_{\nu j} \;  (d_i^\mu)^* d_j^\nu$  &  $\sum_{ij=\{x,y\}} (d_i^i)^* d_j^j$ \quad \\
 \hline
 $\sum_{\mu \nu i j=\{x,y\}} \; \delta_{\mu j} \delta_{\nu i} \; (d_i^\mu)^* d_j^\nu$   &   $\sum_{ij=\{x,y\}} (d_i^j)^* d_j^i $ \quad \\
 \hline
 $\sum_{\mu \nu i j=\{x,y\}} \; \delta_{\mu \nu} \delta_{i j} \;  (d_i^\mu)^* d_j^\nu$  &   $\sum_{ij=\{x,y\}} (d_i^j)^* d_i^j$ \\
 \hline
 $\sum_{\mu \nu i j=\{x,y\}} \;  \delta_{\mu i} \delta_{i j} \delta_{j \nu} \;  (d_i^\mu)^* d_j^\nu$   &  $|d_x^x|^2+ |d_y^y|^2$  \quad\\
\hline
\hline
  \end{tabular}
  \end{adjustbox}
  \label{tab:contraction}
\end{table}
With $\mu=\nu=z$ the only possible contraction is
\begin{gather}
\sum_{i j=\{ x,y\} } \delta_{ij} \;(d_i^z)^* d_j^z= |d_x^z|^2 + |d_y^z|^2.  \label{eq:contractionzz}
\end{gather}

Linear combinations of the four terms from Table~\ref{tab:contraction} and the one in Eq.~\eqref{eq:contractionzz} gives the five terms
in Table~\ref{tab:SOCcomplete} of the main text.
The above SOC induced free energy terms can be also rewritten in terms of the order parameter matrix $\hat{\Delta}$.
Rewriting the five terms in Table~\ref{tab:SOCcomplete} using Eq.~\eqref{eq:OPdef} and linearly recombining them
gives the five independent terms in Table~\ref{tab:SOCcomplete2}, 
from which we see that order parameter products other than $\hat{\Delta}^\dagger \hat{\Delta}$, such as $\hat{\Delta}\sigma_i \hat{\Delta}$
and $\hat{\Delta}\sigma_i \hat{\Delta} \sigma_j$, also appear in the free energy expansion, due to the broken pseudospin $SU(2)$ symmetry~\cite{Vollhardt1990}.
\begin{table}[htp]
  \centering
  \caption{All possible SOC induced GL free energy terms at quadratic order in $\hat{\Delta}$ for the pseudospin triplet pairing states in a 2D model. $\hat{\Delta}_x$ is the part of the order parameter matrix
  $\hat{\Delta}$ that transforms like $k_x$ under the spatial $D_{4h}$ group. The trace Tr is performed in the pseudospin space. Note that, in this table, $f^{\mathrm{SOC}, E}_2$ is allowed because
  the pseudospin Pauli matrix $\sigma_z$ is even under the $xy$-plane mirror reflection $\mathcal{M}_{xy}^{\hat{S}}$. }
  \begin{adjustbox}{width=\linewidth}
      \begin{tabular}{@{\hskip 0.2in} c @{\hskip 0.2in} c  @{\hskip 0.2in}}
  \hline
  \hline
  GL terms & Expressions in terms of $\hat{\Delta}$ \\
  \hline
  $f^{\mathrm{SOC}, A}_2$   &   $ \mathrm{Tr} [\hat{\Delta}^\dagger_x \sigma_z \hat{\Delta}_x \sigma_z ]
+ \mathrm{Tr}  [ \hat{\Delta}^\dagger_y \sigma_z \hat{\Delta}_y \sigma_z ] $     \\
  \hline
   $f^{\mathrm{SOC}, B}_2$   &   $\mathrm{Tr}  [ \hat{\Delta}^\dagger_y \sigma_x \hat{\Delta}_y \sigma_x ]
- \mathrm{Tr}  [ \hat{\Delta}^\dagger_x \sigma_y \hat{\Delta}_x \sigma_y ] $     \\
  \hline
  $f^{\mathrm{SOC}, C}_2$  &   $ \mathrm{Tr}  [ \hat{\Delta}^\dagger_x \sigma_x \hat{\Delta}_x \sigma_x ]
- \mathrm{Tr}  [ \hat{\Delta}^\dagger_y \sigma_y \hat{\Delta}_y \sigma_y ] $     \\
  \hline
  $f^{\mathrm{SOC}, D}_2$  &  $ \mathrm{Tr} [ \hat{\Delta}^\dagger_x \sigma_x \hat{\Delta}_y \sigma_y ]
- \mathrm{Tr}  [ \hat{\Delta}^\dagger_y \sigma_y \hat{\Delta}_x \sigma_x ]  $      \\
  \hline
   $f^{\mathrm{SOC}, E}_2$   &   $i \, \big\{  \mathrm{Tr}  [ \hat{\Delta}^\dagger_x \sigma_z \hat{\Delta}_y ] 
- \mathrm{Tr}  [ \hat{\Delta}^\dagger_y \sigma_z \hat{\Delta}_x  ] \big\}$    \\
\hline
\hline
  \end{tabular}
   \end{adjustbox}
  \label{tab:SOCcomplete2}
\end{table}

\section{SOC dependence of the splitting between helical and chiral states when $J=J^\prime=0$} \label{app:etaJeq0}
\begin{figure}[htp]
\centering
\includegraphics[width=\linewidth,trim={22mm 0mm 0mm 0mm},clip]{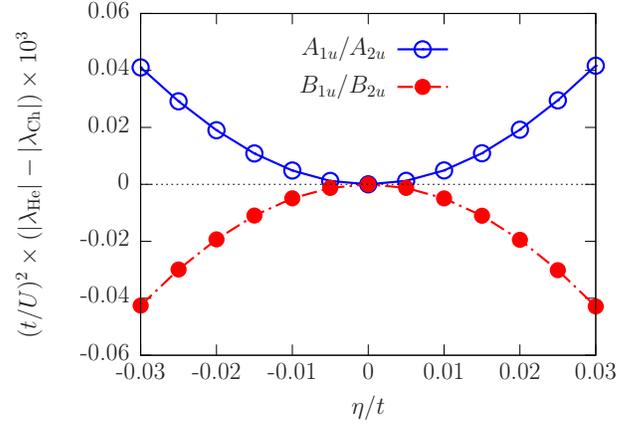}
\caption{$\eta/t$ dependence of the splitting between helical and chiral p-wave pairing states in weak-coupling RG. $t^{\prime\prime\prime} = 0.1t$ and $J=0$.
The splittings are $\propto (\eta/t)^2$ to leading order in $\eta$.}
\label{fig:split_etaJ0}
\end{figure}

The $\eta$ dependence of $\delta \mathcal{F}$ for the 2D model at $J=J^\prime=0$ is calculated in weak-coupling RG and shown In Fig.~\ref{fig:split_etaJ0}. As mentioned in the main text,
we find that the splitting between helical states has a quadratic dependence on $\eta$ for small $\eta$
at $J/U=0$. This result can be understood as follows. When both $\eta=0$ and $J^\prime=J=0$, the two-particle effective interaction
has no coupling between $\alpha+\beta$ bands, which consist of the $d_{xz}$ and $d_{yz}$ orbitals, and the $\gamma$ band, if only intra-band pairing is considered
as in the weak-coupling RG~\cite{Scaffidi2014}. Then the pairing lives purely on the $\gamma$ band since that band has a larger density of states.
Therefore, at zero order in $\eta$, all the off-diagonal density matrices in Eq.~\eqref{eq:deltaFtppp2} are identically zero.
As a consequence, $\delta \mathcal{F} \sim \mathcal{O}(\eta^2)$. However, in general, we expect the three bands to be coupled even when $\eta=0$ if the pair hopping 
$J^\prime \ne 0$. In that case, $\delta \mathcal{F}$ picks up a linear in $\eta$ term, as seen in Fig.~\ref{fig:split_etaJ2}.
The $\eta$ linear term is likely to dominate over the $\eta^2$ term since its estimated $J/U$ for $\mathrm{Sr_2RuO_4}$ is about $0.1$
(see Ref.\onlinecite{Scaffidi2014} and the references therein). Note that even a small $J/U$ can strongly couple the three bands together such that
order parameter magnitudes on the three bands are comparable~\cite{Scaffidi2014}. This is largely because the normal state density of states of the $\alpha +\beta$ bands
is comparable to that of the $\gamma$ band.

\section{SOC induced GL free energy terms at quadratic order for 3D models} \label{app:SOC3D}
For a 3D model that depends on $k_z$, in addition to the basis functions given in Table~\ref{tab:D4h}, an out-of-plane pairing component, with the basis function $\vec{d}(\vec{k})= \hat{z} k_z$ for the $A_{1u}$ and $\vec{d}(\vec{k})=k_z \, (\hat{x}, \hat{y})$ for the $E_u$ representation, is also allowed by symmetry.~\cite{Sigrist1991, Huang2018, Pustogow2019} In the presence of SOC and inter-layer coupling, in general, the vector $\vec{d}(\vec{k})$ of the $E_u$ (but not the $A_{1u}$; see below) representation is a mixture of the in-plane and out-of-plane pairing components, which leads to more GL free energy terms at quadratic order in $\hat{\Delta}$. 

To obtain these GL free energy terms, we follow the 2D derivations
outlined in Appendix~\ref{app:SOCDerivation}. The only difference is that, for 3D models, the $xy$-plane mirror reflection, $\mathcal{M}_{xy}^{\hat{L}+\hat{S}}$, now operates on both the $\vec{k}$ and pseudospin. Besides the terms in Table~\ref{tab:SOCcomplete}, we also get
\begin{subequations}
\begin{align}
f_2^{\mathrm{SOC,6}} & = |d_z^z|^2, \\
f_2^{\mathrm{SOC,7}} & = |d_z^x|^2 + |d_z^y|^2  , \\
f_2^{\mathrm{SOC,8}} & =\sum_{j=\{x,y\}} \big[ (d^z_j)^* d_z^j + c.c.  \big] , \\
f_2^{\mathrm{SOC,9}} & = \sum_{j=\{x,y\}} i \,   \big[ (d^z_j)^* d_z^j - c.c.   \big].
\end{align}
\end{subequations}
$f_2^{\mathrm{SOC,6}}$ and $f_2^{\mathrm{SOC,7}}$ describe the free energy contributions from the $\vec{d}(\vec{k})= \hat{z} k_z$ and $\vec{d}(\vec{k})=k_z \, (\hat{x}, \hat{y})$
pairings, respectively. They exist even without SOC. 
However, $f_2^{\mathrm{SOC,6}}$ and $f_2^{\mathrm{SOC,7}}$ are irrelevant to our current discussions since the $T_c$ of the out-of-plane pairings are expected to be much smaller than that of the in-plane components for $\mathrm{Sr_2RuO_4}$, which is highly quasi-2D. 
The other two GL terms, $f_2^{\mathrm{SOC,8}}$ and $f_2^{\mathrm{SOC,9}}$, describe the coupling between the in-plane $\vec{d}(\vec{k})=\hat{z}\, (k_x, k_y)$ and out-of-plane $\vec{d}(\vec{k})=k_z \, (\hat{x}, \hat{y})$ components within the same $E_u$ representation. Their appearance requires finite SOC to break the full psedospin rotational symmetry. Note that such a coupling does not exist for the $A_{1u}$ representation. 

The mixing of the out-of-plane component to the in-plane chiral p-wave pairing state in $E_u$ leads to a shift of the $T_c$ away from its  zero SOC value, $T_c^0$. On the other hand, the helical p-wave states are unaffected by the mixing; therefore,  the degeneracy among different p-wave states is in general lifted due to the shift. Hence it is important to understand the magnitude of  this shift, and compare it to the effect of other SOC induced GL terms on $T_c$ that we have discussed in Sec.~\ref{sec:microscopic} of the main text. 

To that end, we first analyze the dependence of the GL coefficients associated with $f_2^{\mathrm{SOC,8}}$ and $f_2^{\mathrm{SOC,9}}$, $a_8$ and $a_9$, on small parameters of
the model that we consider~\cite{Roising2019}. Spin rotation symmetry breaking requires that, to leading order in $\eta$, $\{a_8,a_9\} \propto \eta/t $.  Since the terms in $f_2^{\mathrm{SOC,8}}$ and $f_2^{\mathrm{SOC,9}}$ transform like $k_x k_z$ or $k_y k_z$ under spatial rotations, the two GL coefficients necessarily come from a $\vec{k}$ space average $\langle  \cdots k_x k_z \rangle $ or $\langle  \cdots k_y k_z \rangle$, which is nonzero only if the  $k_x k_z$ or $k_y k_z$ dependence is compensated by inter-layer hopping terms such as $t_{yz,xy} \sin k_x/2 \sin k_z/2 \cos k_y/2$ or $t_{xz,xy} \sin k_y/2 \sin k_z/2 \cos k_x/2$. 
As a consequence, to leading order in $t_{yz,xy}/t$ and $t_{xz,xy}/t$, $\{a_8,a_9\} \propto t_z/t$, where $t_z =  \max\{|t_{yz,xy}|,|t_{xz,xy}|\}$. 
Combing the $\eta/t$ and $t_z/t$ dependence, we have $\{a_8,a_9\} \propto  T_c^0 (\eta \, t_z )/t^2 $,  where $T_c^0$ is a characteristic pairing temperature scale for zero SOC.

Now consider the $T_c$ shift of the chiral p-wave pairing state due to the mixing. As mentioned above, we expect that, for $\mathrm{Sr_2RuO_4}$, the $T_c$ of the in-plane and out-of-plane pairing components in the $E_u$ representation, $T_{c, in}$ and  $T_{c, out}$, satisfy $T_{c, in} \gg T_{c, out}$. So the $T_c$ shift
of the in-plane chiral p-wave pairing state due to the mixing
can be well estimated from a second order non-degenerate perturbation theory, i. e. $\delta T_c \approx \max\{ |a_8|^2, |a_9|^2\}/ (T_{c, in} - T_{c, out})$,
which leads to $\delta T_c/T_c \propto (\eta \, t_z )^2 /t^4$.  This shift is negligible compared with the contributions from other SOC induced terms that we have discussed in the main text, since $\eta/t \sim 0.1$ and $|t_z/t| \lesssim 0.1$ for $\mathrm{Sr_2RuO_4}$. 
\end{appendix}

%

%

\end{document}